\documentclass[aps,pra,superscriptaddress,twocolumn,showpacs]{revtex4}

\usepackage{amsmath}
\usepackage{amssymb}
\usepackage{graphicx}
\usepackage{dcolumn}
\usepackage{bm}

\begin{document}

\title{Measurement of phase fluctuations of Bose-Einstein condensates in
an optical lattice}

\author{Bing Wang}
\affiliation{State Key Laboratory of Magnetic Resonance and Atomic
and Molecular Physics, Wuhan Institute of Physics and Mathematics,
Chinese Academy of Sciences, Wuhan 430071, P. R. China}
\affiliation{Key Laboratory of Atomic Frequency Standards (KLAFS),
Wuhan Institute of Physics and Mathematics, Chinese Academy of
Sciences} \affiliation{Graduate School of the Chinese Academy of
Sciences, P. R. China}

\author{Qiang Zhu}
\affiliation{State Key Laboratory of Magnetic Resonance and Atomic and
Molecular Physics, Wuhan Institute of Physics and Mathematics, Chinese Academy
of Sciences, Wuhan 430071, P. R. China}
\affiliation{Graduate School of the
Chinese Academy of Sciences, P. R. China}

\author{Hailong Zhou}
\affiliation{State Key Laboratory of Magnetic Resonance and Atomic
and Molecular Physics, Wuhan Institute of Physics and Mathematics,
Chinese Academy of Sciences, Wuhan 430071, P. R. China}
\affiliation{Key Laboratory of Atomic Frequency Standards (KLAFS),
Wuhan Institute of Physics and Mathematics, Chinese Academy of
Sciences} \affiliation{Graduate School of the Chinese Academy of
Sciences, P. R. China}

\author{Dezhi Xiong}
\affiliation{State Key Laboratory of Magnetic Resonance and Atomic
and Molecular Physics, Wuhan Institute of Physics and Mathematics,
Chinese Academy of Sciences, Wuhan 430071, P. R. China}
\affiliation{Key Laboratory of Atomic Frequency Standards (KLAFS),
Wuhan Institute of Physics and Mathematics, Chinese Academy of
Sciences}

\author{Hongwei Xiong}
\affiliation{State Key Laboratory of Magnetic Resonance and Atomic
and Molecular Physics, Wuhan Institute of Physics and Mathematics,
Chinese Academy of Sciences, Wuhan 430071, P. R. China}

\author{Baolong L\"{u}}
\email{baolong_lu@wipm.ac.cn}
\affiliation{State Key Laboratory of
Magnetic Resonance and Atomic and Molecular Physics, Wuhan Institute
of Physics and Mathematics, Chinese Academy of Sciences, Wuhan
430071, P. R. China} \affiliation{Key Laboratory of Atomic Frequency
Standards (KLAFS), Wuhan Institute of Physics and Mathematics,
Chinese Academy of Sciences}

\date{\today }

\begin{abstract}

Even at zero temperature, there exist phase fluctuations associated
with an array of Bose-Einstein condensates confined in a
one-dimensional optical lattice. We demonstrate a method to measure
the phase fluctuations based on the Fourier spectrum of the atomic
density for a condensate released from the optical lattice. The
phase variance is extracted from the relative intensities of
different peaks in the Fourier spectrum. This method works even for
high lattice strength where interference peaks disappear in the
atomic density distribution.

\end{abstract}

\begin{pacs}
  {03.75.Lm, 67.85.Hj, 37.10.Jk, 67.10.Ba}
\end{pacs}

\maketitle

\section{INTRODUCTION}
\label{Introduction}

For Bose-Einstein condensates in an optical lattice, the phase
fluctuation is a significant quantity in the investigation of
quantum phase transitions \cite{Bloch}. The transition from
superfluid to Mott insulator is usually accompanied by significantly
increased phase fluctuations which can be manifested in the
interference pattern of the condensate samples. The vanishing of the
contrast of the interference fringes is widely regarded as a
characteristic of the quantum phase transition.

The simplest lattice configuration suitable for demonstrating phase
fluctuations is a one-dimensional (1D) standing-wave laser field
loaded by Bose-Einstein condensates (BEC). Such a 1D optical lattice
is actually a common tool to test quantum properties of the cold
atoms in periodic potentials. It has been used to demonstrate phase
coherence \cite{Pedri,ZhouXiaoji}, Bloch oscillations
\cite{BlochOscillation}, number squeezed state \cite{Kasevich},
Josephson current \cite{JosephsonCurrent}, nonlinear self-trapping
of matter waves \cite{NLSelftrapping,NLST-Wubiao}, and so on. In
theory, quantum fluctuations in phase and atomic number are often
illustrated by considering a BEC in a double-well potential
\cite{TwoMode-1,TwoMode-2,TwoMode-3,TwoMode-4,TwoMode-5,Junction-1,Junction-2,Junction-3}.
Of course, experimental measurement of the phase fluctuation plays a
key role in understanding the quantum process occurring in a
lattice.

In a pioneering work by Orzel \emph{et al.} \cite{Kasevich}, phase
fluctuations of the subcondensates in a 1D optical lattice were
measured by using the interference pattern of the released
condensates. The phase variance was extracted from the contrast of
the observed interference peaks. However, at very high lattice
depth, the typical interference peaks disappear completely due to
the large phase fluctuations, and this method is thus not valid any
more. In Ref. \cite{LatticeDepth}, it is also shown that, close to
the Mott insulator, the vanishing of the interference fringes makes
it difficult to describe the quantitative changes of the system
controlled with further increased lattice depth.

In this paper, we develop a method to measure the phase variance by
employing the Fourier spectrum of the released atomic cloud.
Particularly, a simple analytical expression is found to extract the
phase fluctuations. Our method works in principle even when the
visibility of the interference peaks is completely lost, as
demonstrated in our experiment. It is expected that this method
provides a unique tool to other phase transitions in cold atomic
systems \cite{RMPBloch}.

The paper is organized as follows. In Sec.\ \ref{Theory}, we present
the theoretical model to extract the phase fluctuations from the
Fourier spectrum of the atomic density for a condensate released
from the optical lattice. In Sec.\ \ref{Experiment}, the theoretical
model is demonstrated in our experiment. The accuracy and validity
of our method are discussed, and a comparison with the method in
\cite{Kasevich} is also presented. Finally in Sec.\
\ref{Conclusion}, we summarize our obtained results.

\section{THEORY}
\label{Theory}

We now consider a 1D optical lattice in the dimension of the $x$
axis, formed by a conventional standing-wave laser field. Its
strength is usually measured in units of the recoil energy
$E_{r}=h^{2}/2m\lambda^{2}$, where $m$ is the atomic mass, $h$ is
the Planck constant, and $\lambda$ the optical wavelength. In the
tight-binding limit, the condensate loaded to the lattice can be
treated as a chain of disk-shaped subcondensates equally spaced by
the lattice period $d=\lambda/2$. The total number of the lattice
sites occupied by the condensate is denoted by $M$. When suddenly
released from the optical lattice, the condensate undergoes a free
expansion process. After a time of flight (TOF) of $\tau$, the wave
function of the atomic cloud can be written as
\[
\Psi(x,\tau)=\sum_{l=1}^{M}\alpha_{l}\Phi_{l}(x,\tau),
\]
where $\Phi_{l}(x,\tau)$ refers to the wave function of the
subcondensate initially centered at the $i$th lattice site, and
$\left|\alpha_{l}\right|^{2}$ represents the probability for an atom
roughly located at the $l$th lattice site. The atomic density is
then written as
\begin{equation}\label{eq:Density}
\begin{split}
|\Psi(x,\tau)|^{2} & = \sum_{l,q}\alpha_{l}^{*}\alpha_{q}\Phi_{l}^{*}(x,\tau)\Phi_{q}(x,\tau)\\
 & =G_{0}+\sum_{n=1}^{M-1}G_{n},
\end{split}
\end{equation}
where $G_{0}$ reads
\[
G_{0}=\sum_{l=1}^{M}\left|\alpha_{l}\right|^{2}\left|\Phi_{l}(x,\tau)\right|^{2},
\]
 while $G_{n}$ with $n\geq1$ takes the following form:
\begin{equation}\label{eq:Gn}
G_{n}=\sum_{l=1}^{M-n}\alpha_{l}^{*}\alpha_{l+n}\Phi_{l}^{*}(x,\tau)\Phi_{l+n}(x,\tau)
+\text{c.c.}.
\end{equation}

Note that, $G_{0}$ is physically different from other $G_{n}$ with
$n \geq 1$. It contains no interference terms, and is just a direct
sum of the atomic densities of all the subcondensates. Therefore, it
gives rise to a spatially smooth density profile. In contrast,
$G_{n}$ describes the interference between the subcondensates spaced
by $nd$ in the optical lattice. One characteristic associated with
$G_{n}$ must be mentioned. The integration of $G_{n}$ is zero ($\int
G_{n}\text{d}x=0$) due to the orthogonality between different
$\Phi_{l}$, which implies that $G_{n}$ would give rise to
interference structures rather than a smooth background in the
density distribution.

\subsection{Fourier spectrum}

According to Eq.\,\eqref{eq:Gn}, the Fourier transform of $G_{n}$ is
\begin{equation}\label{eq:Fn}
\begin{split}
F_{n} & = \frac{1}{\sqrt{2\pi}}\sum_{l}\alpha_{l}^{*}\alpha_{l+n}\int\Phi_{l}^{*}(x,\tau)\Phi_{l+n}(x,\tau)e^{ikx}\text{\ensuremath{\text{d}x}}+\text{c.c.}\\
 & =\frac{1}{\sqrt{2\pi}}\widetilde{w}_{n}(k,\tau)\sum_{l}\alpha_{l}^{*}\alpha_{l+n}e^{ikld}+\text{c.c.}.
\end{split}
\end{equation}
Here,
\begin{equation}\label{eq:Correlation}
\widetilde{w}_{n}(k,\tau)=\int\Phi^{*}(x,\tau)\Phi(x-nd,\tau)e^{ikx}\text{\ensuremath{\text{d}x}}
\end{equation}
 is independent of the lattice site $l$. The second line of
 Eq.\,\eqref{eq:Fn} is obtained using the fact that all $\Phi_{l}$ are identical wave functions except for their
center positions. In the tight-binding limit the Wannier function
$\Phi(x,t=0)$ can be well approximated by a Gaussian wave packet
$(\pi\sigma^{2})^{-1/4}\exp(-x^{2}/2\sigma^{2})$, where
$\sigma=\sqrt{\hbar/m\widetilde{\omega}_{x}}$ is the oscillator
length, $m$ the atomic mass and $\widetilde{\omega}_{x}/2\pi$ the
axial trapping frequency of the lattice wells. After the TOF, the
expanded subcondensate has a much larger size than its initial wave
packet ($\hbar\tau/m\gg\sigma^{2}$), then the wave function of a
single subcondenste can be written as \cite{Baolong}
\begin{equation}\label{eq:Expanded wavepacket}
\Phi(x,\tau) =
\frac{1}{\pi^{1/4}\sigma^{1/2}}\left(1+\frac{i\hbar\tau}{m\sigma^{2}}\right)^{-1/2}\exp\left(\frac{imx^{2}}{2\hbar\tau}\right).
\end{equation}

Substituting Eq.\,\eqref{eq:Expanded wavepacket} into
Eq.\,\eqref{eq:Correlation}, one gets
\begin{equation}\label{eq:wn-peaks}
\widetilde{w}_{n}(k,\tau)
 =W_{n}\int\exp\left[ix\left(k-nk_{1}\right)\right]\text{\ensuremath{\text{d}x}},
\end{equation}
where $k_{1}=2\pi/\lambda_{1}$, $\lambda_{1}=2\pi\hbar\tau/dm$ and
\[
W_{n}
 =\frac{1}{\sqrt{\pi}\sigma}\left|1+\frac{i\hbar\tau}{m\sigma^{2}}\right|^{-1}\exp
 \left(\frac{in^{2}md^{2}}{2\hbar\tau}\right).
\]
Here, $\lambda_{1}$ is a characteristic length, equal to the travel
distance of an atom with a velocity twice the single photon recoil
velocity. The integration term yields a $\delta$-function like peak
at $k=nk_{1}$, and the peak width is inversely proportional to the
spatial size of the expanded wave packets. Similar analysis to the
conjugate part in $F_{n}$ yields an identical peak at the symmetric
position, $k=-nk_{1}$, instead. Therefore, the Fourier transform of
$G_{n}$ shows a pair of peaks at $k=\pm nk_{1}$ (only one peak for
$G_{0}$). Apparently, the characteristic length $\lambda_{1}$ is
actually the spatial period of the Fourier component corresponding
to the peak of the $n=1$ order. As such, coherence properties
associated with different spacings between subcondensates in the
optical lattice can be distinguished from one another just by
inspecting the Fourier spectrum of the density distribution of the
expanded atomic cloud. The whole power spectrum of the atomic
density is simply given by $S(k)=\Sigma_{n}\left|F_{n}\right|^{2}$.

Note that, $\lambda_{1}$ is much larger than the initial condensate
size, which means $kld\ll1$, and hence $e^{ikld}\simeq 1$. One may
get from Eq.\,\eqref{eq:Fn} the peak intensities in the power
spectrum $S(k)$:
\begin{equation}\label{eq:Fn-2}
P_{n}=\left|F_{n}(k=nk_{1})\right|^{2}=AY_{n},
\end{equation}
where $A=\left|\widetilde{w}_{0}(k=0)\right|^{2}/2\pi$ and
$Y_{n}=\left|\sum_{l=1}^{M-n}\alpha_{l}^{*}\alpha_{l+n}\right|^{2}$.
In particular, $Y_{0}=1$, as required by the normalization
condition. From the expression in Eq.\,\eqref{eq:wn-peaks}, one sees
that the amplitude of $\widetilde{w}_{n}(k=nk_{1})$, and hence $A$,
is independent of $n$. Therefore, the relative intensity of peaks in
the power spectrum depends only upon $Y_{n}$.

\subsection{Phase Fluctuations}
\label{Subsec-PhaseFluctuations}

We now turn to consider the peak intensities in the power spectrum,
from which the phase fluctuation can be  deduced. In the optical
lattice, the confined subcondensates undergo phase fluctuations. The
phase factor of each subcondensate is contained in the corresponding
coefficient $\alpha_{l}$, and the summation term $Y_{n}$ in
Eq.\,\eqref{eq:Fn-2} is then rewritten as
\begin{equation*}
\begin{split}
Y_{n} & =\left|\sum_{l=1}^{M-n}\left|\alpha_{l}\alpha_{l+n}\right|e^{i\phi_{ln}}\right|^2\\
 & \simeq \left(\sum_{l}\left|\alpha_{l}\alpha_{l+n}\right|\right)^2
 \left|\frac{1}{M-n}\sum_{l}e^{i\delta\phi_{ln}}\right|^2,
\end{split}
\end{equation*}
where $\delta\phi_{ln}\equiv\phi_{l+n}-\phi_{l}$. Due to phase
fluctuations of the subcondensates when trapped in the optical
lattice, $\delta\phi_{ln}$ takes random values with zero average.
 The last line of the above equation is obtained by assuming that $\left|\alpha_{l}\alpha_{l+n}\right|$
changes slowly with $l$. As long as $n\ll M$, the summation term
$\sum_{l}\left|\alpha_{l}\alpha_{l+n}\right|$ is constant for
different $n$, i.e.,
$\sum_{l}\left|\alpha_{l}\alpha_{l+n}\right|\simeq\sum_{l}\left|\alpha_{l}\right|^{2}=1$.
$Y_{n}$ is then simplified to:
\begin{equation}\label{eq:Yn}
Y_{n} = \left|\frac{1}{M-n}\sum_{l}e^{i\delta\phi_{ln}}\right|^2.
\end{equation}
\begin{figure}[h]
\centering
\includegraphics[width=0.9\columnwidth,angle=0]{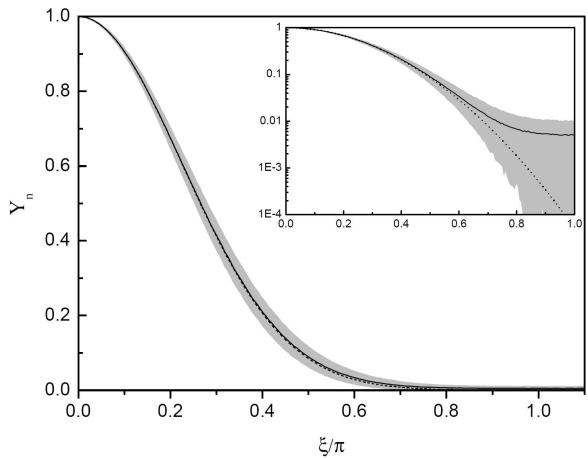}
\caption{The solid line shows the numerically calculated $Y_{n}$
versus the variance of the random phase $\delta\phi_{ln}$ for a
total lattice number $M=200$. Each data point is an average of
$1000$ individual
 runs of the computation. Shaded area corresponds to
the standard error. The dashed line is an exponential curve in the
form of $e^{-\xi^2}$. The inset is the same figure on a logarithm
vertical scale, highlighting the discrepancy between the two
curves.} \label{Fig1}
\end{figure}

Apparently, for $n>0$, $Y_{n}$ depends on the variance of
$\delta\phi_{ln}$, denoted by $\xi^2=<\delta\phi_{ln}^2>$. As shown
in Fig.\,\ref{Fig1}, $Y_{n}$ drops quickly with increasing $\xi$.
The randomness of $\delta\phi_{ln}$ results in the fluctuations of
$Y_{n}$ around its averaged values. In the region of larger phase
variance, $Y_{n}$ has larger fractional fluctuations. In
computation, the total lattice site number is assumed to be $M=200$.
We find, however, the averaged value of $Y_{n}$ is nearly unchanged
for different values of $M$.

Checking the summation $\sum_{l}e^{i\delta\phi_{ln}}$ in
Eq.\,\eqref{eq:Yn}, one sees that it is nearly a real number as the
imaginary terms are averaged to zero. By replacing $\delta\phi_{ln}$
by its variance $\xi$ and assuming $\xi \ll 1$, this summation term
can be approximated by $(M-n)e^{-\frac{1}{2}\xi^2}$. We then have an
approximated expression of $Y_{n}$ in an exponential form:
\begin{equation}\label{eq:Yn-1}
Y_{n}=e^{-\xi^2}.
\end{equation}
This exponential curve is plotted in Fig.\,\ref{Fig1} as well.
Surprisingly, it shows a good match to the numerical data (the solid
curve in Fig.\,\ref{Fig1}) even for the region far beyond $\xi \ll
1$. For example, the relative error is only $5\%$ at $\xi=0.5\pi$.
Of course, the tendency to larger relative errors can be clearly
seen as $\xi$ is increased.

The phase fluctuation of two adjacent subcondensates can be measured
by a phase variance $\sigma^{2}=\langle \delta\phi_{l1}^{2}\rangle$.
It is straightforward to prove that two subcondensates spaced by $n$
times the lattice period correspond to a $n$ times larger phase
variance, i.e., $\xi^2 =n\sigma^{2}$. One knows from
Eq.\,\eqref{eq:Fn-2} and \eqref{eq:Yn-1} that
$P_{n}=Ae^{-n\sigma^{2}}$. Taking natural logarithm for both sides
of this formula, one obtains
\begin{equation}\label{eq:Peak-phase}
\ln P_{n}=\ln A-n\sigma^{2},
\end{equation}
 which clearly shows the linear relation between the logarithmic scale
of the peak height and subcondensate spacing $n$. The slope of this
linear curve is just the phase variance $\sigma^{2}$. Therefore, the
phase fluctuations of the condensates confined in discrete lattice
wells can be easily determined by the peak structures in the Fourier
spectrum of the atomic density distribution after the time of
flight. We stress here that this method to measure $\sigma^{2}$ does
not need exact calibration of the peak intensities, because
$\sigma^{2}$ is unaffected by any identical scale factors applied to
all $P_{n}$. This clearly shows the convenience of Eq.\
\eqref{eq:Peak-phase} in the measurement of $\sigma^{2}$.

There also exists an alternative method to measure the phase
variance, where $\sigma^{2}$ is determined by comparing the
experimentally measured $P_{n}$ with numerically calculated $Y_{n}$.
When $\xi$ is very large, the analytical expression of $Y_{n}$ in
Eq.\,\eqref{eq:Yn-1} is not a good approximation any more. In this
case, the second method is more reliable than the former one based
on Eq.\,\eqref{eq:Peak-phase}. We shall make a comparison between
the two methods later.

\section{EXPERIMENT}
\label{Experiment}

Our experiments are carried out by using a nearly pure $^{87}$Rb
condensate in the hyperfine state $|F=2, M_{F}=2\rangle$ with
typically $10^{5}$ atoms. The experimental setup was described
elsewhere \cite{Baolong}. The 1D optical lattice is formed by a
retroreflected laser beam with a wavelength of
$\lambda=1064\,\text{nm}$. Its strength is calibrated using a method
of Kapitza-Dirac scattering \cite{KapitzaDirac-1,LatticeDepth}. The
recoil energy is $E_{r}=h\times 2.03\,\text{kHz}$. The lattice light
is adiabatically applied to the cigar-shaped condensate along its
axial direction during a time of $50\,\text{ms}$. After a holding
time of $10\,\text{ms}$, the lattice light, as well as the magnetic
trap, are suddenly switched off. Finally, an absorption image is
taken for the released atomic cloud after a $30\,\text{ms}$ of TOF,
by using a probe light directed perpendicular to the lattice beam.
The experimental parameters correspond to a characteristic length
$\lambda_{1}=259\,\mu\text{m}$, which is much larger than the pixel
size ($9\,\mu\text{m}$) of our CCD camera. In principle, the peaks
up to the order of $n=14$ can be resolved by the CCD camera. In
order to obtain the statistics of the phase variance, the experiment
was repeated at least eleven times for each lattice depth.
\begin{figure}[h]
\centering
\includegraphics[width=0.9\columnwidth,angle=0]{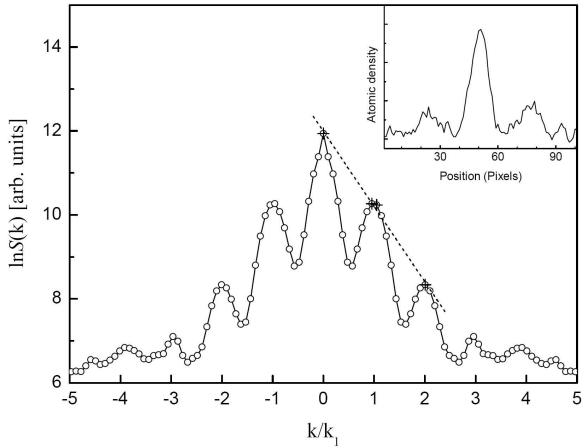}
\caption{Open circles show the power spectral intensity obtained
from the density distribution of an expanded condensate released
from an optical lattice with a depth of $34.6\,E_{r}$. Four points
at the top of three peaks are marked by crosses, and used to
determine a phase variance of $\sigma^{2}=1.79(8)$ by a linear fit
(dashed line). The peak close to $k/k_{1}=3$ is caused by optical
noises of the probe light. The inset shows the atomic density
distribution of this atomic cloud along the direction of the lattice
beam, showing clearly two side peaks due to the interference of the
subcondensates.}
 \label{Fig2}
\end{figure}

The power spectrum $S(k)$ is obtained from an absorption image as
follows: Spatial frequency spectrum is calculated by
Fourier-transforming the matrix of image. Taking the absolute square
to obtain the 2D power spectral density, and $S(k)$ is then obtained
from the 2D matrix by making summation along the direction
perpendicular to the lattice beam. Thus, only the power spectrum
along the lattice direction is remained in $S(k)$. Finally, taking
the natural logarithm of $S(k)$, we can find the peaks and extract
$\sigma^{2}$ according to Eq.\,\eqref{eq:Peak-phase}.

Figure\,\ref{Fig2} displays a power spectrum corresponding to the
absorption image of an atomic cloud released from the optical
lattice with a depth of $34.6\,E_{r}$. As predicted, $S(k)$ consists
of a series of peaks equally spaced by $k_{1}$. The peaks with
$n=0\text{--}2$ are true signals of the cold atoms, whereas the peak
close to $k/k_{1}=3$ is confirmed to be optical noise of the probe
light itself. This noise peak appears occasionally in repeated
experiments, even in the absence of the atomic cloud.

A higher phase variance means a weaker phase correlation. In our
experiment, subcondensate pairs with a spacing larger than $2d$ can
not yield a peak ($n>2$) high enough to be visible in $S(k)$. We
thus inferred that the phase correlation of two subcondensates drops
quickly with increased distance between them, which is a natural
consequence of the proportional relation of $\xi^{2}=n\sigma^{2}$.
\begin{figure}[h]
\centering
\includegraphics[width=0.9\columnwidth,angle=0]{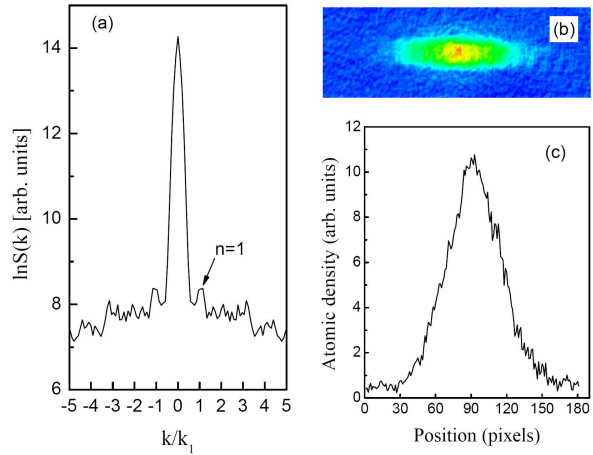}
\caption{(Color online) (a) Power spectrum $S(k)$ of an expanded
condensate initially trapped at a depth of $68\,E_{r}$. (b) A
false-color absorption image of this released condensate, with a
field of view of $0.55\ \text{mm}\times 1.63\ \text{mm}$. (c) Atomic
density in the $x$ direction, which is obtained by integrating the
pixels in height.} \label{Fig3}
\end{figure}

To extract the phase variance $\sigma^2$, we used the data points at
the top of peaks in a power spectrum to perform the linear fit in
the form of Eq.\, \eqref{eq:Peak-phase}. The standard error of a
fitted slope ($\sigma^2$) is usually small ($\leq 5\%$). In
contrast, the value of $\sigma^2$ shows much larger fluctuations
from shot to shot as the experiment is repeated under the same
conditions. We thus take only the latter fluctuations into
consideration when calculating the error bars of the phase variance.
A linear fit to the three peaks in Fig.\,\ref{Fig2} yields a phase
variance of $\sigma^{2}\simeq 1.79$. At the corresponding depth
level of the optical lattice, the atomic density profile exhibits
clear interference peaks along the lattice direction (see the inset
in Fig.\,\ref{Fig2}). As the lattice depth is increased, the phase
variance is increased, and the higher order peaks in $S(k)$ with
$n\ge 1$ are expected to become weaker accordingly.
Figure\,\ref{Fig3} displays a typical result in such a case. At the
depth of $68\,E_{r}$, only the peaks of $n=0,1$ are observed. Higher
order peaks are too weak to be identified. As seen from
Fig.\,\ref{Fig3} (b) and (c), the released condensate has lost the
interference structures completely, in contrast to the side peaks in
Fig.\,\ref{Fig2}. In fact, despite the loss of interference peaks in
atomic density distribution, the peaks in $S(k)$ can still be
observed for strong optical lattice, up to the highest level
 ($120\,E_{r}$) we have reached.

\begin{figure}[h]
\centering
\includegraphics[width=0.9\columnwidth,angle=0]{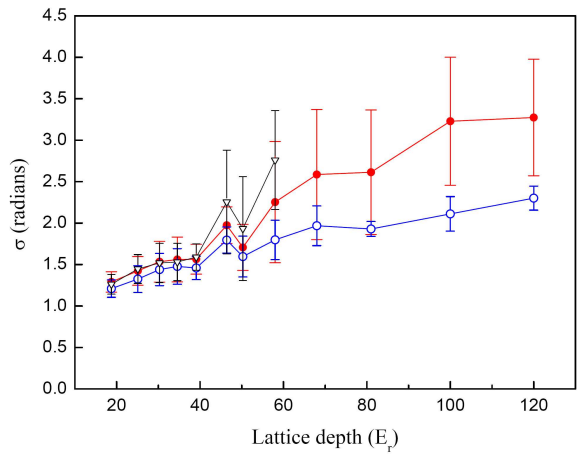}
\caption{(Color online) Measured phase variance $\sigma$ versus the
lattice depth. For each data point, eleven independent runs were
averaged. Blue data points were obtained based on
Eq.\,\eqref{eq:Peak-phase}, whereas, the red data points were
obtained by comparing the measured relative intensity of the peaks
of $n=1$ with the numerical values of $Y_{1}$. Black data points
were obtained by fitting the atomic density profile using the method
in \cite{Kasevich}.}
 \label{Fig4}
\end{figure}

The dependence of phase variance $\sigma$ upon the lattice depth is
displayed in Fig.\,\ref{Fig4}. The two sets of data (blue and red
ones) were obtained using the two methods described in
Sec.\,\ref{Subsec-PhaseFluctuations}. There exists a clear trend,
where the deeper the optical lattice, the larger the phase variance.
The value of $\sigma$ grows from $\sim 0.4\pi$ to $\sim \pi$,
indicating that the relative phase between adjacent lattice sites
gains increased randomness, and tend to be a completely random
phase. Below $50\,E_{r}$, the blue data points are very close to the
red ones. Beyond this level, however, the discrepancy between the
two sets of data becomes larger with increased lattice depth. It can
be simply understood by the fact that the exponential form of
$Y_{n}$ (Eq.\,\eqref{eq:Yn-1}) is not a good approximation for large
$\xi$. Roughly speaking, the linear fit method can only be applied
when $\xi$ is less than $0.6\pi$, as is evident in Fig.\,\ref{Fig4}.
It is worthy to point out that, although the linear fit method is
invalid in this case, the alternative method is still simple to
extract the phase variance from the Fourier spectrum.

The accuracy of phase variance $\sigma$ is limited by at least two
factors. First, as shown by the solid curve in Fig.\,\ref{Fig1},
$Y_{n}$ is less sensitive to $\xi$ when $\xi$ becomes large,
particularly in a region around $\pi$. Second, $Y_{n}$ itself is in
fact a fluctuating parameter in principle, although we have taken
its averaged value as the measured result. The fluctuation is
evident from the fact that the relative intensity of each peak in
$S(k)$ varies from shot to shot. Therefore, in the deep lattice
region, data points of $\sigma$ are always accompanied by large
error bars.

As a comparison, we have also extracted the phase variance by
analyzing the visibility of the interference peaks of the released
atomic cloud. This method was first demonstrated in \cite{Kasevich},
where the visibility is characterized by a quantity $\zeta$ defined
as the ratio of the width of a single peak to separation between the
peaks. The phase variance is determined by comparing the observed
value of $\zeta$ with those obtained from simulated data sets.
First, we fit the interference profile with three Gaussian peaks to
get the value of $\zeta$. Then, we calculate the interference
profile by using a simple one-dimension model which is similar to
that in our previous work \cite{Baolong}. Each subcondensate in a
single lattice well is assigned a random phase. Those random phases
are set in such a way that the phase difference between two adjacent
lattice sites obeys a Gaussian distribution with a given variance
$\sigma^{2}$. We convolve the calculated interference profile with a
resolution function to account for the limited resolution of our
imaging system. To obtain the simulated $\zeta$, we fit the
convolved waveform with the same fit function applied to the
experimental data. The simulation procedure is repeated many times
to obtain an averaged value of the simulated $\zeta$.

The phase variances deduced from this fitting method are displayed
in figure 4. It is obvious that, below a depth level of $\sim
60\,E_{r}$, the results of the fitting method agree well with our
method. This further confirms the validity of our method. However,
above $60\,E_{r}$, the fitting method does not work due to the
following reasons: To strictly follow the method demonstrated in
\cite{Kasevich}, both the peak width and peak separation must be
treated as fitting parameters. However, at high lattice-depth level,
only one broad peak is left in the interference profile, as shown in
Fig.\,\ref{Fig3}(c). The fitting procedure cannot give reasonable
peak positions. More specifically, the side peaks obtained from the
fitting program deviate significantly from the positions where they
should be located. In addition, the fitted width of the side peaks
is not reliable.

\section{SUMMARY AND DISCUSSION}
\label{Conclusion}

We have developed a method to measure the phase fluctuations of the
subcondensates confined in a 1D optical lattice. In our method, the
Fourier spectra of the conventional absorption images of the
released atomic clouds have been investigated. The phase variance
between adjacent lattice wells is deduced from the relative
intensities of the peaks in a Fourier power spectrum. Our
experimental measurements have displayed an increased phase variance
as the lattice depth becomes larger, and also indicated that phase
correlation of two lattice wells decreases quickly with the
increased distance between them. Our method does not rely on the
existence of interference peaks of the released condensates, and it
works even for very large lattice depth. This method is a
complementary to that demonstrated in \cite{Kasevich}, and will be a
useful tool in analyzing phenomenons associated to phase
fluctuations in optical lattice systems, particularly for the case
close to the quantum phase transition.

Our theoretical model is established in the tight-binding limit,
where a condensate in the optical lattice can be regarded as a chain
of subcondensates. We found that, for weak optical lattice, there is
no multi-peak structure in $S(k)$. The peaks start to appear when
the lattice depth reaches a level of $\sim 10\,E_{r}$. Well resolved
peaks can be observed if the depth level is further increased to
$\sim 20\,E_{r}$. It sets a coarse boundary beyond which our model
is applicable. Below the level of $20\,E_{r}$, one has to switch to
the method in \cite{Kasevich} to measure the phase variance.

The optical lattice is not homogeneous due to the presence of the
magnetic trap which is used to support the atoms against the
gravity. The harmonic confinement of the magnetic trap corresponds
to a trapping frequency of $2\pi\times 7.6\ \text{Hz}$, and it
remains until the sudden release of the atomic cloud. For a total
atomic number of $10^{5}$, the number of lattice sites that are
populated is $M\simeq 200$. Most atoms are distributed in the center
region of the lattice where the tunneling rate $J$ is nearly
uniform. This assures the assumption that $\sigma$ is uniform over
the optical lattice. On the other hand, we did not see noticeable
changes of $\sigma$ as the total atomic number is changed from
$4-15\times 10^{4}$. It is due to the fact that the tunneling rate
is independent of the atom numbers in single lattice wells.

In principle, our method can be extended to 2D and 3D optical
lattices by treating the power spectrum in each dimension
separately. For a 2D optical lattice, one probe beam perpendicular
to the lattice plane is enough. For a 3D lattice, however, an
additional probe beam is required to detect the atomic density
profile in the third dimension.

\begin{acknowledgments}
This work is supported by the National Natural Science Foundation of
China under Grant Nos. 11174331, 11175246 and 11104322, and by the
National Key Basic Research and Development Program of China under
Grant No. 2011CB921503.
\end{acknowledgments}

\end{document}